\documentclass[twocolumn,showpacs,preprintnumbers,amsmath,amssymb,aps]{revtex4}
\usepackage{graphicx}
\usepackage{amsthm}
\usepackage{epsfig}
\usepackage{mathrsfs}
\usepackage[hypertex,linkcolor=red]{hyperref}
\def \vett #1{\boldsymbol #1}
\def\Tr{\operatorname{Tr}}
\def\<{\langle}\def\>{\rangle}
\def\spc #1{\mathscr #1}
\def\map #1{\mathcal #1}  
\def\Cmplx{\mathbb C} 
\def\SU #1{\mathbb{SU} (#1)}
\def\SO #1{\mathbb{SO} (#1)}
\def\spc #1{\mathscr #1}
\begin{document}
\title{Superbroadcasting and classical information}

\author{Giulio Chiribella} 
\author{Giacomo M. D'Ariano} 
\author{Chiara Macchiavello} 
\author{Paolo Perinotti}
\affiliation{Dipartimento di Fisica ``A. Volta'' and CNISM, via Bassi
  6, I-27100 Pavia, Italy.}
\author{Francesco Buscemi} \affiliation{ERATO-SORST Quantum
  Computation and Information Project, Japan Science and Technology
  Agency, Daini Hongo White Bldg. 201 5-28-3, Hongo, Bunkyo-ku Tokyo
  113-0033 Japan}

\date{\today}

\begin{abstract}
  We address the problem of broadcasting $N$ copies of a generic qubit
  state to $M>N$ copies by estimating its direction and preparing a
  suitable output state according to the outcome of the estimate. This
  semiclassical broadcasting protocol is more restrictive than a
  general one, since it requires an intermediate step where classical
  information is extracted and processed. However, we prove that a
  suboptimal superbroadcasting, namely broadcasting with simultaneous
  purification of the local output states with respect to the input
  ones, is possible. We show that in the asymptotic limit of
  $M\to\infty$ the purification rate converges to the optimal one,
  proving the conjecture that optimal broadcasting and state
  estimation are asymptotically equivalent. We also show that it is
  possible to achieve superbroadcasting with simultaneous inversion of
  the Bloch vector direction (universal NOT). We prove that in this
  case the semiclassical procedure of state estimation and preparation
  turns out to be optimal. We finally analyse semiclassical
  superbroadcasting in the phase-covariant case.
\end{abstract}


\maketitle
\section{Introduction}
A basic feature of classical information is that it can be copied and
distributed to an unlimited number of users. However, as one considers
quantum information the fundamental task of broadcasting for pure states is
impossible, and this implies severe limitations to very useful
purposes such as parallel computation, networked communications, and
secret sharing.\par

A perfect distribution of the information encoded into
$N$ input systems equally prepared in a pure state to $M>N$ users  would
correspond to the so called \emph{quantum cloning}, which is forbidden
by the laws of quantum mechanics \cite{clon}. Nevertheless, the case of
mixed input states is different, since one needs only
the local state of each final user to be equal to the input state,
whereas the global output state is allowed to be correlated. This fact
opens the possibility to generalize the idea of cloning  
to quantum maps that output correlated
states such as their local reduced states are copies of the input.
This generalized version of quantum cloning was named {\em quantum
  broadcasting} in Ref.~\cite{nobro}. In this work the impossibility 
of perfect broadcasting was
proved in the case of a single input copy whenever the set of states
to be broadcast contains a pair of noncommuting density matrices.
This proof was later often considered as the mixed states-scenario extension of
the no-cloning theorem. However, it was recently
shown  that even noncommuting quantum
states can be perfectly broadcast provided a suitable number of input
copies is available \cite{Broad1}.  
Moreover, a new phenomenon can occur,
which was named \emph{superbroadcasting}: for $N$ two-level
systems (qubits), equally prepared in an unknown mixed input state,
the information contained in the direction of the Bloch vector can be
distributed to $M>N$ users and the local state of each final user can be more
pure than the initial copies.

An intuitive explanation of the superbroadcasting effect is provided by the statement that
superbroadcasting shifts the noise from local purities to global correlations~\cite{Broad1,Broad2}.
One of the issues of superbroadcasting is then a deeper understanding of the role of correlations of
different nature. While there are correlations which improve the accessibility of information
encoded in multiple systems~\cite{rafal}, the case of superbroadcasting points out that other kind
of correlations are in fact detrimental in this respect.  This leads to an amount of information in
the global output state that is lower than the sum of informations contained in the local reduced
states, i.~e. the total information in absence of correlations.  Natural questions then arise at this
stage.  Are the correlations among the final users solely quantum, or is it possible to purify the
local states by introducing just classical correlations?  Moreover, in the optimal broadcasting
protocol, the distribution of information is achieved by coherently manipulating input systems, and
the true direction of the Bloch vector remains unknown during the whole procedure. What happens if
one first uses the $N$ input copies to estimate the direction of the Bloch vector, and then
distributes $M$ pure states pointing in the estimated direction?  Is it still possible, on average,
to increase the purity of local states?

A preliminary extensive analysis of bipartite correlations at the output of superbroadcasting maps
suggests that no bipartite entanglement is present~\cite{Broad2}, whereas the analysis of
multipartite entanglement is still an open problem. The fact that the practical protocol for
achieving superbroadcasting involves pure state cloning~\cite{CEM,Broad3} suggests on the other hand
that the output state contains quantum correlations coming from the structure of the tensor product
Hilbert space and its symmetric subspace.

In this paper, we will consider a semiclassical procedure for broadcasting, which consists of
measurement and subsequent repreparation of the quantum states, usually referred to as the
\emph{measure-and-prepare} scheme. We call this scheme semiclassical because broadcasting occurs via
extraction and processing of classical information, though the information is retrieved by a
collective measurement which might be strictly quantum, being generally a nonlocal measurement. We
show that the phenomenon of superbroadcasting can still be observed in this case, even though the
scaling factors obtained by this scheme are suboptimal. Such a procedure introduces only classical
correlations among the final copies, as the joint output state remains fully separable. The
remarkable presence of superbroadcasting even in the semiclassical scenario can be explained as a
change of encoding of the classical information about the direction of the Bloch vector. In fact,
the tensor product of $N$ identical qubit states provides an encoding of direction, in which the
information is spread in a nonlocal way over the whole $N$-qubits system. In order to extract such
an information, one needs either a collective measurement or a statistical processing of
single-qubit measurements.  However, after the information has been extracted, it can be
redistributed exploiting a new encoding, which is more favourable to single users.  This result can
be interpreted as a proof that on one hand optimal superbroadcasting involves quantum effects that
cannot be simulated by extracting and re-using classical information, and on the other hand the
phenomenon of superbroadcasting itself is improved by entanglement but not necessarily due to it.
Moreover, we will show that the fidelity of the optimal estimation of direction coincides with the
fidelity of the optimal superbroadcasting protocol in the limit $M \to \infty$. This provides the
first example of generalization to arbitrary mixed states of the relation between cloning and state
estimation, which was known in the literature for pure states \cite{bruss-ek-macc,bruss-macc,acin}.

In this paper we will also address the optimal approximation of a
\emph{universal NOT broadcasting}, namely the impossible
transformation which corresponds to a combination of ideal
purification, quantum cloning, and spin flip (universal NOT). We will
derive the optimal physical map, observing how in this case the
semiclassical procedure achieves the optimal fidelity. In other words,
the optimal universal NOT broadcasting can be viewed as a purely classical
processing of information, as it happens in the case of pure input
states~\cite{UNot}.

The paper is organised as follows. In Sect. \ref{sec:preliminary} we 
introduce the main tools that will be employed to describe symmetric and 
covariant 
broadcasting maps. In Sect. \ref{sec:estimation} we derive the covariant 
superbroadcasting map achieved by optimal estimation of the direction of the 
Bloch vector and conditional repreparation of the $M$ output states. In Sect. \ref{sec:not} we derive 
the optimal covariant NOT broadcasting map and show that it can be achieved 
by semiclassical means. 
In Sect. \ref{sec:phase} we study the phase covariant case, we derive the 
phase covariant semiclassical map and compare it with the universal case.
Finally, in Sect. \ref{sec:conc} we summarise the main results of this paper 
and discuss their perspectives.

\section{Preliminary tools}\label{sec:preliminary}

\subsection{Schur-Weyl duality and permutation invariant operators}

Symmetry considerations play a fundamental role in the analysis of
broadcasting maps, where the input states are $N$ identically prepared
states, and the output states are required to be permutationally
invariant, in order to equally distribute information among many
users. A very convenient tool to deal with permutation invariance is
the so-called Schur-Weyl duality, which relates the irreducible
representations of the permutation group to the irreducible
representations of the group $\SU d$.  

For a system of $N$ qubits, it is possible to decompose the Hilbert
space $\spc H= (\Cmplx^2)^{\otimes N}$ as a Clebsch-Gordan direct sum
\begin{equation}\label{SpaceDecomp}
  \spc H \simeq \bigoplus_{j = j_0}^{N/2}~ \spc H_j \otimes \Cmplx^{m_j}~,
\end{equation}
where $j_0$ is 0 (1/2) for $N$ even (odd), $\spc H_j= \Cmplx^{2j +1}$, and 
\begin{equation}\label{mj}
m_j= \frac{2j+1}{M/2 +j+1} \binom{M}{M/2 -j}~.
\end{equation}  
Here $j$ is the quantum number associated to the total angular
momentum, and the spaces $\spc H_j$ carry the irreducible
representations of $\SU 2$. In other words, for any $U_g \in \SU 2$,
we have
\begin{equation}\label{WedSU2}
  U_g^{\otimes N}= \bigoplus_{j=j_0}^{N/2} U_g^{(j)} \otimes \openone_{m_{j}}~,
\end{equation}
where $\{U_g^{(j)}\}$ is the irreducible representation labeled by
the quantum number $j$, and $\openone_{m_j}$ is the identity in
$\Cmplx^{m_{j}}$.

According to the Schur-Weyl duality, the action of a permutation of
the Hilbert spaces in the tensor product $\spc H^{\otimes N}$ can be
represented in the same way as in Eq. (\ref{WedSU2}), with the only
difference that the roles of $\spc H_j$ and $\Cmplx^{m_j}$ are
exchanged, namely the action of permutation is irreducible in
$\Cmplx^{m_j}$ and is trivial in $\spc H_j$.  In this decomposition, a
permutation invariant operator $X$ has the form
\begin{equation}\label{XInv}
X = \bigoplus_{j=j_0}^{N/2}  X_j \otimes \openone_{m_j} ~.
\end{equation}
In particular, the state $\rho^{\otimes N}$ of $N$ identically
prepared qubits can be written as
\begin{equation}\label{CEM}
\rho^{\otimes N} = \bigoplus_{j=j_0}^{N/2}~ \rho_j  \otimes \frac{\openone_{m_j}}{m_j}~,
\end{equation} 
where $\rho_j$ is a positive operator on the Hilbert space $\spc H_j$
with $\sum_{j=j_0}^{N/2}\Tr[\rho_j]=1$. This decomposition, and the
explicit expression for the $\rho_j$'s was first given in
\cite{CEM}.  If the single-qubit state is $\rho= (\openone + r \vett n
\cdot \vett \sigma)/2$, where $r$ and $\vett n$ are the length and the
direction of the Bloch vector, respectively, then $\rho_j$ is given by
\begin{equation}\label{RhoJ}
\begin{split}
  \rho_j &=m_j~ (r_+ r_-)^{N/2}\left(\frac{r_+}{r_-}\right)^{\vett n\cdot\vett J^{(j)}}\\
  &= m_j~ (r_+ r_-)^{N/2} \sum_{m=-j}^{j} \left( \frac{r_+}{r_-}
  \right)^m ~ |j,m;\vett n\>\<j,m;\vett n|~,
\end{split}
\end{equation}
where $|j,m;\vett n\>$ is the eigenstate of the operator $\vett n
\cdot \vett J^{(j)}$ for eigenvalue $m$, and $r_{\pm}= (1 \pm r)/2$.

\subsection{Symmetric broadcasting maps}

In order to derive the optimal universal NOT broadcasting we will make
use of the formalism of the Choi isomorphism between CP
maps $\map E$ from states on the Hilbert space $\spc H$ to states on
the Hilbert space $\spc K$, and positive operators $R$ on $\spc H
\otimes \spc K$.  The isomorphism is given by
\begin{equation}\label{R}
  R_{\map E} = \map E \otimes \mathcal{I} (|\Omega\>\<\Omega|) \longleftrightarrow \map E (\rho) = \Tr[ \openone \otimes \rho^T R_\mathcal{E}]~,
\end{equation} 
where $|\Omega\> \in \spc H \otimes \spc H$ is the non normalized
maximally entangled vector $|\Omega\> =\sum_m |m\>|m\>$, and $X^T$
denotes the transpose of $X$ with respect to the fixed basis
$\{|m\>\}$.  For a broadcasting map from $N$ input qubits to $M$
output qubits, we have $\spc H = (\Cmplx^2)^{\otimes N}$ and $\spc K=
(\Cmplx^2)^{\otimes M}$.

In the study of universal broadcasting maps, one requires the
universal covariance property, which ensures that the output Bloch
vectors point at the same direction as the input ones, and is defined
as follows
\begin{equation}\label{CovMap}
\map E (U_g^{\otimes N} \rho U_g^{\otimes N \dag}) = U_g^{\otimes M} \map E (\rho) U_g^{\otimes M \dag}~, 
\end{equation}
$\rho$ being any state on $\spc H = (\Cmplx^2)^{\otimes N}$, and $U_g$
being any element of $\SU 2$. The universal covariance of the map
$\map E$ translates into the commutation relation
\begin{equation}\label{CovSU2}
  \left[ R, U_g^{\otimes M} \otimes U_g^{*\otimes N}\right]=0 \qquad \forall U_g \in \SU 2~.
\end{equation}
Using the property $U_g^*= \sigma_y U_g \sigma_y$, this relation can
be rewritten as
\begin{equation}\label{CovSU2'}
[S, U_g^{\otimes M +N}]=0 \qquad \forall U_g \in \SU 2~,
\end{equation}
where 
\begin{equation}\label{S}
  S= (\openone^{\otimes M} \otimes \sigma_y^{\otimes N})~R~  (\openone^{\otimes M} \otimes \sigma_y^{\otimes N})~.
\end{equation}

Moreover, since the figures of merit for broadcasting maps are usually
averaged over the output states, without loss of
generality we can consider maps that are invariant under permutations of
the output systems. Similarly, since we consider only permutation
invariant input states, we can restrict attention to maps which are
invariant under permutation of the input systems. Consequently $R$ can
be required to satisfy
\begin{equation}\label{CovPerm}
[R, \Pi^M_{\sigma} \otimes \Pi^N_{\tau}]=0\qquad \forall \sigma \in S_M, \forall \tau \in S_N~,
\end{equation}
where $\sigma$ $(\tau)$ are permutations of the $N$ input ($M$ output)
qubits, and $\Pi_{\sigma}^N$ ($\Pi_{\tau}^M$) are the unitary
operators representing them.  Clearly, this relation is equivalent to
\begin{equation}\label{CovPerm'}
[S, \Pi_{\sigma}^{M} \otimes \Pi_{\tau}^{N}]=0~.
\end{equation}
Exploiting the decomposition (\ref{SpaceDecomp}), we can write
\begin{equation}
\spc H \otimes \spc K = \left( \bigoplus_{j = j_0}^{M/2} \spc H_j \otimes \Cmplx^{m_j} \right) \otimes \left(\bigoplus_{l=l_0}^{N/2} \spc H_l \otimes \Cmplx^{m_l} \right)~, 
\end{equation}  
and, rearranging the factors in the tensor product, we have
\begin{equation}
\spc H\otimes \spc K = \bigoplus_{j=j_0}^{M/2} \bigoplus_{l=l_0}^{N/2} (\spc H_j \otimes \spc H_l) \otimes (\Cmplx^{m_j} \otimes \Cmplx^{m_l})~.
\end{equation}
According to Eq. (\ref{CovPerm'}), the operator $S$ must be invariant under
permutations of both the input and the output qubits, whence it must
have the form (\ref{XInv})
\begin{equation}
S= \bigoplus_{j=j_0}^{M/2}  \bigoplus_{l=l_0}^{N/2} S_{jl} \otimes (\openone_{m_j} \otimes \openone_{m_l})~, 
\end{equation}
where $S_{jl}$ is a positive operator on $\spc H_j \otimes \spc H_l$.
Moreover, the product $\spc H_j \otimes \spc H_l$ can be further
decomposed as
\begin{equation}
\spc H_j \otimes \spc H_l= \bigoplus_{J=|j-l|}^{j+l}  \spc H_J^{j,l}~,
\end{equation}
where $\spc H_J^{j,l}$ are the $(2J+1)$-dimensional subspaces that
carry the irreducible representations of the Clebsch-Gordan series of
$U_g^{(j)} \otimes U_g^{(l)}$. According to Eq.~(\ref{CovSU2'}), $S$
must be invariant under $U_g^{\otimes(N+M)}$, therefore
\begin{equation}
S= \bigoplus_{j=j_0}^{M/2} \bigoplus_{l=l_0}^{N/2} \bigoplus_{J=|j-l|}^{j+l}  s^J_{j,l} ~P^J_{j,l} \otimes \openone_{m_j} \otimes \openone_{m_l}~,
\end{equation}   
where $P_{j,l}^J$ is the projection from $\spc H_j \otimes \spc H_l$
onto $\spc H_J^{j,l}$, and $s^J_{j,l}$ are positive reals.

To find the optimal broadcasting maps, it is useful to know the extremal
points of the convex set of the corresponding operators.  According to
the classification given in Ref.~\cite{Broad2}, a map is extremal if
and only if
\begin{equation}
S= \bigoplus_{l=l_0}^{N/2} \frac{2l +1}{2 J_l +1}\frac{1}{m_{j_l}} ~P^{J_l}_{j_l,l} \otimes \openone_{m_{j_l}} \otimes \openone_{m_l}~, 
\label{ExtS}
\end{equation}
where $J_l$ and $j_l$ are two vectors of quantum numbers functions of $l$ (of course, the entries of
$j_l$ can range from $j_0$ to $M/2$, and while the entries of $J_l$ range from $|j_l-l|$ to $j_l
+l$. For universally covariant superbroadcasting one has $J_l=|l-M/2|$ and $j_l=M/2$ \cite{Broad1}).

\section{Superbroadcasting via optimal estimation of direction}
\label{sec:estimation}

Let us consider a broadcasting map that distributes to $M$ users the
information contained into $N$ qubits, each of them prepared in the
same unknown state
\begin{equation}\label{InState}
\rho (\vett n,r) = \frac{1}{2} \left ( \openone + r \vett n \cdot \vett \sigma \right)~, 
\end{equation} 
$r$ and $\vett n$ being the length and the direction of the Bloch
vector, respectively. Precisely, with the term ``information'' we mean
the information about the direction $\vett n$, while the degree of
mixedness of the input state is regarded only as an effect of noise.
Accordingly, the aim of the broadcasting procedure is to distribute to
each user a local state with a Bloch vector pointing in a direction as
close as possible to the direction $\vett n$, and possibly with higher
purity.

Here we want to obtain the broadcasting map in two steps, namely by
first performing a measurement on the initial qubits, in order to
optimally extract the classical information about their direction, and
then by preparing $M$ identical pure states pointing in the estimated
direction. This approach can also be used for the NOT broadcasting,
with the only difference that after estimation we have to prepare pure
states pointing in the opposite direction.

In the following we denote with $\hat{ \vett n}$ the estimated
direction of the Bloch vector, and the measurement statistics will be
described by a positive operator valued measure (POVM) $M(\hat {\vett
  n})$, namely by a set of positive semidefinite operators satisfying
the normalization condition
\begin{equation}
  \int_{\mathbb S^2} d^2 \vett n~ M(\vett n)=\openone,
\end{equation}
where $d^2 \vett n$ is the normalized Haar measure on the unit
sphere $\mathbb S^2$.  The probability density of estimating
$\hat{\vett n}$ when the true direction is $\vett n$ is given by the
Born rule $p(\hat{\vett n}|\vett n)=\Tr[M(\hat{\vett n}) \rho (\vett
n,r)]$. Once the estimation is performed the output state of the
broadcasting procedure is
\begin{equation}\label{OutState}
\rho^M_{out}(\vett n, r) = \int_{\mathbb S^2}~ d \hat{\vett n}~ p(\hat {\vett n}|\vett n) ~|\hat{\vett n}\> \<\hat {\vett n}|^{\otimes M}~,
\end{equation} 
where $|\vett n \>$ denotes the eigenvector of $\vett n\cdot \vett
\sigma$ for the eigenvalue $+1$ [a NOT broadcasting can be obtained
replacing $|\vett n\>$ with its orthogonal complement $|-\vett
n\>$ in the above formula]. Accordingly, the local state of
each user is
\begin{equation} 
\rho_{out}^1(\vett n,r)= \int d \hat{\vett n}~ p(\hat{\vett n}|\vett
n)~|\hat{\vett n}\>\<\hat {\vett n}|~,
\end{equation}
and it is independent of the number of users $M$.

In the following we will require the broadcasting map to be covariant
under rotations. This corresponds to require the property
\begin{equation}\label{CovBroad}
\rho_{out}^M(g \vett n,r)= U_g^{\otimes M}~ \rho^M_{out}(\vett n,r) U_g^{\otimes M \dag}~, 
\end{equation}  
where $g \in \SO 3$ denotes a rotation in the three-dimensional space, and $U_g \in \SU 2$ is a two
by two matrix representing the rotation $g$ in the single-qubit Hilbert space. In other words, we
require that, if the Bloch vector of the input copies is rotated by $g$, then also the output state
is rotated by the same rotation.  In order to have a covariant broadcasting map the POVM density
$M(\vett n)$ must be itself covariant, namely it must satisfy the property\cite{Holevo}
\begin{equation}
M(g ~ \vett n) = U_g^{\otimes N}~ M(\vett n)~ U_g^{\otimes N~ \dag}~,
\end{equation}
for any rotation $g$. In this way, the probability distribution has
the property
\begin{equation}\label{InvProb}
p(g \hat{\vett n}|g \vett n)=p(\hat{\vett n}|\vett n) \qquad \forall g \in \SO 3~,
\end{equation} 
and, therefore, the output state (\ref{OutState}) satisfies the
covariance property (\ref{CovBroad}).

In this framework, we want the local state $\rho^1_{out}(\vett n,r)$
to be as close as possible to the pure state $|\vett n\>\<\vett n|$.
For this purpose, the estimation strategy will be optimised in order to
maximize the single-site fidelity
\begin{eqnarray}\label{Fid}
F\left( \rho^1_{out}(\vett n,r), |\vett n\>\<\vett n|\right)&=& \int_{\mathbb S^2} d \hat{\vett n}~ p(\hat{\vett{n}}|\vett n)~ |\<\hat{\vett n}|\vett n\>|^2\\&=& \int_{\mathbb S^2} d \hat{\vett n}~ p(\hat {\vett n}|\vett n)~ \frac{1 + \hat{\vett n}\cdot \vett n}{2}~. 
\end{eqnarray}  
In the case of the universal NOT broadcasting, the single-user output state is
\begin{equation}
\tilde \rho_{out}^1(\vett n,r) = \int_{\mathbb{S}^2} d \hat{\vett n}~ p(\hat{\vett n}| \vett n)~|-\hat{\vett n}\>\<-\hat{\vett n}|~,
\end{equation}
and one considers its fidelity with the pure state\break $|-\vett n\>\<-\vett n|$.  Clearly, in the
classical procedure both broadcasting and NOT broadcasting have the \emph{same fidelity}.  Due to
the invariance property (\ref{InvProb}), the fidelity does not depend on the actual value of the
direction $\vett n$, and it is enough to maximize it for a fixed direction, for example the positive
direction $\vett k$ of the $z$-axis. For this reason, from now on we will denote the fidelity simply
with $F$.

The estimation strategy that maximizes the fidelity $F$ can be found
in a simple way by
exploiting the decomposition~(\ref{CEM}) of the input state. First,
due to the special form of the states, without loss of generality
we can restrict our attention to POVMs of the form
\begin{equation}\label{BlockPovm}
M(\vett n)= \bigoplus_{j=j_0}^{N/2}~ M_j (\vett n) \otimes
\openone_{m_j}~, \end{equation} 
where each $M_j(\vett n)$ is a POVM in the representation space $\spc
H_j$, namely $M_j(\vett n)\ge 0$ and 
\begin{equation}
\int d \vett n~ M_j(\vett n)= \openone_{2j+1}~.
\end{equation}
In fact, if $\widetilde M(\vett n)$ is any POVM, then the
corresponding probability distribution is
\begin{equation}
\begin{split}
  p(\hat{\vett n}|\vett n)&= \sum_j~ \Tr\left[\widetilde M(\hat{\vett
      n})~
    \left(\rho_j(\vett n,r) \otimes \frac{\openone_{m_j}}{m_j}\right)\right]= \nonumber\\
  &\sum_j\Tr[ \widetilde M_j(\hat{\vett n})~ \rho_j (\vett n,r)]~,
\end{split}
\end{equation}  
where $\widetilde M_j(\vett n)=1/m_j~\Tr_{m_j}[\widetilde M(\vett
n)]$.  The same probability distribution can be obtained by a POVM of
the form (\ref{BlockPovm}), just by choosing $M_j(\vett n)=\widetilde
M_j(\vett n)$.
 
The fidelity~(\ref{Fid}) becomes then a sum of independent contributions $F= \sum_j~ f_j$ with 
\begin{equation}
f_j= \int_{\mathbb S^2} d \hat {\vett n}~\Tr[ M_j(\hat{\vett
n}) \rho_j(\vett k,r)]~ |\<\hat {\vett n}|\vett k\>|^2~,
\end{equation}      
where $\vett k$ is the unit vector pointing in the positive
$z$-direction. Since all contributions are independent, each of them
can be maximized separately. For this purpose, we can exploit the
result by Holevo~\cite{Holevo} about the optimal estimation of
rotations for mixed states. For any value $j$ the optimal POVM is
given by
\begin{equation}
M_j(\vett n)= (2j +1) |j,j;\vett n\> \<j,j;\vett n|~,
\end{equation} 
where $|j,j;\vett n\>$ is the eigenvector of $\vett n \cdot \vett
J^{(j)}$ corresponding to the eigenvalue $j$, and the contribution to
the fidelity is
\begin{equation}
f_j(\vett n)= \frac{1}{2} \left(~ 1 + \frac{\Tr[\rho_j(\vett k,r) J_z^{(j)}]}{j+1}~\right)~.
\end{equation}
Finally, by using the expression (\ref{RhoJ}), we can calculate
explicitly the fidelity as
\begin{equation}\label{OptFidEst}
F= \frac{1}{2} \left[ ~ 1 +  (r_+ r_-)^{N/2} \sum_{j=j_0}^{N/2}~
\frac{m_j}{(j+1)}  \sum_{m=-j}^{j} m
\left(\frac{r_+}{r_-}\right)^m \right]~.  
\end{equation}
As already mentioned, this is also the value of the fidelity for the
universal NOT broadcasting obtained via optimal estimation of
the direction.

Now we want to investigate whether the phenomenon of superbroadcasting
takes place in the classical broadcasting procedure.
To do this we consider the Bloch vector of
the local output state, by writing
\begin{equation}\label{OutBloch}
\rho^1_{out}(\vett n,r)= \frac{1}{2} \left( \openone + r' (r)\vett n'(\vett n)\cdot \vett \sigma\right)~.  
\end{equation}
The first observation is that the direction $\vett n'=\vett n'(\vett n)$ of the output state is the same as
the direction $\vett n$ of the input state $\vett n$. Due to covariance, it is enough to prove this
fact for $\vett n$ pointing in the $z$-direction. In order to prove it, suppose that $\vett n' \not
= \vett k$, then we would have
\begin{equation}
\<\vett n'| \rho^1_{out}(\vett k,r) |\vett n'\> \ge \<\vett k|\rho_{out}^1(\vett
k,r) |\vett k\>=F~,
\end{equation} 
namely the fidelity of the output state with $|\vett n'\>$ would be
higher than the fidelity with $|\vett k\>$ (the equality holds only if
the output state is maximally mixed). Since we can write $\vett n' =
\bar g \vett k$ for some suitable rotation $\bar g$, in that case we
could replace the optimal POVM $M(\vett n)$ with a new POVM $M'(\vett
n)= M(\bar g^{-1} \vett n)$, where $g^{-1}$ denotes the inverse
rotation of $g$. In this way the fidelity associated to the new POVM
would be $F'= \<\vett n'| \rho_{out}^1(\vett k,r) |\vett n'\> \ge F$,
in contradiction to the fact that $M(\vett n)$ is the optimal POVM.
Therefore, for the optimal POVM the Bloch vector of the output state
\emph{must} point in the same direction as the Bloch vector of the
input state.

\begin{figure}[h]
\epsfig{file=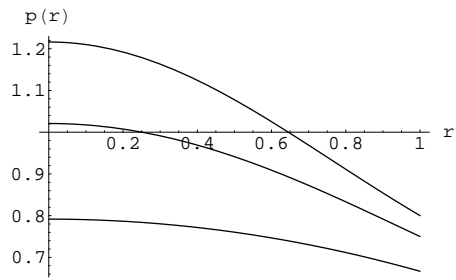,width=6cm}
\caption{Scaling factor $p(r)=r'/r$, with $r'$ given in
  Eq.~(\ref{FinalLenght}), for the classical universal broadcasting
  procedure. The three curves, from bottom to top, refer to $N=4,6,8$
  input copies, respectively. Notice that for $N=4$ there is no
  superbroadcasting, namely we always have $r'<r$. The same plots also
  describe the optimal universal NOT broadcaster described in
  Section~\ref{sec:not}.}\label{f:fig1}
\end{figure}

Once we know that the Bloch vector of the output state points in the
correct direction, we can simply calculate its length $r'$ by the
relation
\begin{equation}
F=\<\vett n| \rho_{out}^1(\vett n,r) |\vett n\> = \frac{1 +r'}{2}~,
\end{equation}
which is straightforward from Eq. (\ref{OutBloch}). 
Thus we obtain
\begin{equation}\label{FinalLenght}
r'(r)=  (r_+ r_-)^{N/2}\sum_{j=j_0}^{N/2}~
\frac{m_j}{(j+1)}  \sum_{m=-j}^{j} m
\left(\frac{r_+}{r_-}\right)^m~. 
\end{equation}
The significant parameter in order to assess the quality of
broadcasting is the {\em scaling factor} given by the ratio of input
and output single site Bloch vector length $p(r)=r'/r$, which is
plotted in Fig.~(\ref{f:fig1}) for $N=4,6,8$.

The plot of this expression demonstrates the presence of superbroadcasting for $N \ge 6$: in this
case the length of the Bloch vector of each single-user state is increased after the broadcasting
map. In other words, the classical broadcasting procedure allows to distribute to many users the
information about the direction of the input Bloch vector, and, at the same time, to increase the
purity of the final local states.  This proves that superbroadcasting can be achieved with a
classical procedure, by first extracting the classical information about the Bloch vector direction,
and then distributing this information among the final users. The increase in the length of the
Bloch vector at each site corresponds to an encoding of information which is more favourable to each
single user.

Moreover, the expression~(\ref{FinalLenght}) can be compared with the
corresponding one for the optimal universal
superbroadcasting~\cite{Broad1,Broad2}, where the information about
the direction is not extracted from the input states, but coherently
manipulated and distributed. Remarkably, in the asymptotic limit of a
large number $M$ of output copies, the two expressions coincide,
namely the optimal distribution of information is achieved
asymptotically by the classical broadcasting procedure. This result
provides the generalization to mixed states of the well known relation
between cloning and state
estimation~\cite{bruss-ek-macc,bruss-macc,acin}.

\section{Optimal universal NOT broadcasting}
\label{sec:not}

As mentioned above, a set of $N$ qubits, equally prepared in the state 
$\rho (\vett n,r)=
(\openone +r \vett n\cdot \vett \sigma)/2$, can be viewed as an
encoding of the classical information about the direction $\vett n$.
Suppose now that we want to distribute such an information to $M>N$
users, and, at the same time, change the encoding by flipping the
direction of the Bloch vector.  In other words, we are interested in
the best approximation of the impossible transformation
\begin{equation}\label{IdealUNot}
 ~\rho (\vett n,r)^{\otimes N} \longrightarrow  |-\vett n\>\<-\vett n|^{\otimes M}~.
\end{equation}
For pure states, and for $N=M=1$, this transformation
coincides with the universal NOT gate, which flips the spin of a qubit
for any possible direction~\cite{UNot}. In general, the
transformation~(\ref{IdealUNot}) corresponds to the combination of a
perfect purification of the input states, followed by a perfect $N
\rightarrow M$ cloning, and by a perfect flipping of the output
qubits. We will call the impossible transformation (\ref{IdealUNot})
\emph{ideal universal NOT broadcasting}.

In order to derive the optimal CP  map $\map N$ approximating the transformation (\ref{IdealUNot}), we define the single-site output state
\begin{equation}
\rho_{out}^1 (\vett n,r) = \Tr_{M-1} \left[~ \map N (\rho (\vett n, r)^{\otimes N}) \right]~,
\end{equation}
where $\Tr_{M-1}$ denotes the partial trace over $M-1$ output qubits.
Notice that, since we consider symmetric broadcasting maps, the output
state must be invariant under permutations, and the above definition
does not depend on the choice of the $M-1$ qubits that are traced out.
The optimization of the map $\map N$ corresponds then to the
maximization of the single-site fidelity
\begin{equation}\label{FidNot}
F (\rho_{out}^1(\vett n, r), |-\vett n\>\<-\vett n|) = \<-\vett n|~\rho_{out}^1(\vett n,r)~|-\vett n\>~.
\end{equation}
Here we consider universal broadcasting, which corresponds to require
the fidelity to have the same value for any direction $\vett n$.
Accordingly, the search for the optimal map can be restricted to the
class of maps with the covariance property~(\ref{CovMap}). In the
following, we denote by $F_\textrm{NOT}$ the value of the single-site
fidelity~(\ref{FidNot}). Moreover, since $F_\textrm{NOT}$ is a linear
function of the map $\map N$, in order to maximize $F_\textrm{NOT}$ we
can restrict the attention to the set of extremal universal
broadcasting maps. Such extremal maps are a finite number, and in the
formalism of the Choi isomorphism~(\ref{R}) are
characterized by Eq.~(\ref{ExtS}).

To evaluate $F_\textrm{NOT}$, we choose the direction $\vett k$ of the
$z$-axis in (\ref{FidNot}), and exploit the relation
\begin{equation}
  r' \doteq \Tr[
  \rho_{out}^1(\vett k,r) \sigma_z]=1-2F_\textrm{NOT}~.
\end{equation}
Therefore the optimal map corresponds to the minimum value for $r'$.
The value of $r'$ for an extremal map has been calculated in
Refs.~\cite{Broad1,Broad2}
\begin{equation}
r'= \frac{2}{M} (r_+ r_-)^{N/2} \sum_{l=l_0}^{N/2} ~ \beta( J_l,j_l,l)~m_l~\sum_{n=-l}^l n \left(\frac{r_-}{r_+}\right)^n,
\end{equation}
where
\begin{equation}
\beta (J,j,l)=\frac{J(J+1)-j(j+1)-l(l+1)}{2l(l+1)}~.
\end{equation}
Since $r_- \le r_+$, the sum $\sum_{n=-l}^l n (r_-/r_+)^n$ is always
negative, and the maximization of the fidelity corresponds to the
maximization of $\beta$ over $J$ and $j$. Of course, for fixed values
of $j$ and $l$, to maximize $\beta$ one has to take $J$ maximum, i.~e.
$J=j+l\equiv J_l$. Moreover, since $\beta (j+l,j,l)= j/(l+1)$, the
maximum $\beta$ is obtained by maximizing also $j$, i.~e. by taking
$j=M/2\equiv j_l$. Therefore $\beta_{\max}= M/(2l+2)$, corresponding
to the fidelity
\begin{equation}
  F_\textrm{NOT}= \frac{1}{2} \left[ 1+(r_+r_-)^{N/2} \sum_{l=l_0}^{N/2} \frac{m_l}{l+1} \sum_{m=-l}^l m \left( \frac{r_+}{r_-} \right)^m\right] ~.
\end{equation}     
Remarkably, this expression coincides with the expression
(\ref{OptFidEst}) of the fidelity of the optimal estimation of
direction.  This proves that there is no better way of performing
universal NOT broadcasting than first estimating the direction of the
Bloch vector, and subsequently preparing the $M$ output qubits in the
opposite direction with respect to the estimated one, analogously to what
happens in the case of pure input states~\cite{UNot,peris}.

\section{Phase covariant case}\label{sec:phase}

In this section we consider the \emph{phase covariant} case, where in Eq.~(\ref{CovMap}), instead of
allowing $U_g$ to move within the whole $\mathbb{SU}(2)$ group, we restrict it to belong to a fixed
proper subgroup $\mathbb{U}(1)\subset\mathbb{SU}(2)$ of rotations around a fixed axis, say around
the $z$-axis.  Channels satisfying this covariance property act ``equally well'' not on the whole
Bloch sphere, as in the universal case, but only on circles orthogonal to the rotation axis.
Intuitively, since the phase covariance property is not as strict as in the universal case, we
expect that phase covariant procedures generally achieve better performances compared to their
universal counterparts.

In Ref. \cite{Broad2} the optimal phase covariant superbroadcasting was derived and was shown to act
more efficiently than the optimal universal superbroadcasting.  Analogously to the procedure of
Section~\ref{sec:estimation} for the universal case, we will now try to figure out whether there
exists a classical procedure that achieves phase covariant superbroadcasting and reaches the
fidelity of the optimal phase covariant superbroadcaster in the limit of infinite number of final
users. In this Section we show that in fact such a measure-and-prepare scheme exists and consists of
an optimal phase estimation over mixed qubit states~\cite{mixphase} followed by the preparation of a
suitable pure state.

Let us start considering input states lying on the $xy$-equator of the
Bloch sphere, namely
\begin{equation}\label{eq:phase-fam}
  \rho(\phi,r)=\frac 12(\openone+r\cos\phi\sigma_x+r\sin\phi\sigma_y).
\end{equation}
We then require
covariance under the one-phase rotations group around the $z$-axis,
namely
\begin{equation}
U_\phi=e^{i\phi\sigma_z/2}.
\end{equation}
The action of a unitary operator $U_\phi$ over a state of the
form~(\ref{eq:phase-fam}) is
\begin{equation}
U_\phi\rho(\phi_0,r)U_\phi^\dag=\rho(\phi_0+\phi,r),
\end{equation}
whence it is clear that the action of $\mathbb{U}(1)$ rotates the
Bloch vector around the $z$-axis without affecting its length, namely without 
changing the purity of the state.

\begin{figure}[h]
\epsfig{file=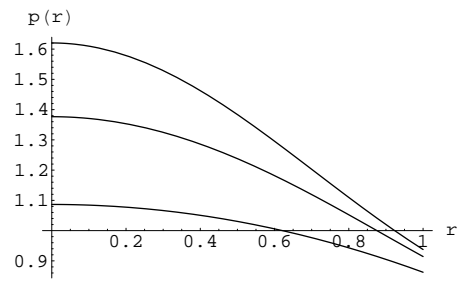,width=6cm}
\caption{Scaling factor $p(r)=r'/r$, with $r'$ given in
  Eq.~(\ref{eq:rphase}), for the classical phase covariant
  broadcasting procedure. The three curves, from bottom to top, refer
  to $N=4,6,8$ input copies, respectively. Compared with the universal
  case, shown in Fig.~\ref{f:fig1}, the phase covariant procedure
  always achieves better performances.}\label{fig2}
\end{figure}

The semiclassical phase covariant broadcasting procedure we propose is
the following. We optimally estimate the value of the phase $\phi$ by
a measurement over $N$ copies of $\rho(\phi,r)$ given by 
the POVM density $P(\phi)$ derived in Ref.~\cite{mixphase}
\begin{equation}
P(\phi)=U_\phi^{\otimes N}\xi(U_\phi^\dag)^{\otimes N}.
\end{equation}
In the above expression the \emph{seed} $\xi$ of the optimal POVM is given by
\begin{equation}
  \xi=\bigoplus_{j=j_0}^{N/2}(2j+1)\sum_{n=-j}^j|j,n;\vett k\>\<j,n;\vett k|\otimes\openone_{m_j},
\end{equation}
wgere $\vett k$ is the rotation axis and $|j,n;\vett k\>$ denotes eigenvectors of the the angular
momentum along $\vett k$ with total angular momentum $j$. The POVM density $P(\phi)$ obeys the
normalization condition
\begin{equation}
\int_0^{2\pi}\frac{d\phi}{2\pi}P(\phi)=\openone.
\end{equation}
After performing the estimation, which gives the conditional probability
density $p(\hat\phi|\phi)=\Tr[\rho(\phi,r)P(\hat\phi)]$,
the output state for $M$ final users is prepared as
\begin{equation}
  \rho^M_{out}(\phi,r)=\int_0^{2\pi}d\hat\phi\ p(\hat\phi|\phi)\ |\hat\phi\>\<\hat\phi|^{\otimes M},
\end{equation}
where $|\phi\>$ denotes the eigenvector of
$\cos\phi\sigma_x+\sin\phi\sigma_y$ for the eigenvalue +1. As in the previous 
sections, we focus on the single-site reduced output, namely
\begin{equation}
\rho^1_{out}(\phi,r)=\int_0^{2\pi}d\hat\phi\ p(\hat\phi|\phi)\ |\hat\phi\>\<\hat\phi|,
\end{equation}
and the fidelity of this procedure is given by
\begin{equation}\label{FidPhase}
F=\<\phi|\rho^1_{out}(\phi,r)|\phi\>.
\end{equation}
Following the same arguments presented in
Section~\ref{sec:estimation}, it can be proved that
$\rho^1_{out}(\phi,r)$ and $\rho(\phi,r)$ have parallel Bloch vectors,
that is,
\begin{equation}
  \rho^1_{out}(\phi,r)=\frac 12(\openone+r'\cos\phi\sigma_x+r'\sin\phi\sigma_y),
\end{equation}
and the fidelity can be again calculated as $F=(1+r')/2$. By exploiting the 
results of
Ref.~\cite{mixphase}, the single-site output Bloch vector length $r'$
turns out to be
\begin{equation}\label{eq:rphase}
  r'=4(r_+r_-)^{N/2}\sum_{j=j_0}^{N/2}m_j\Tr\left[E^{(j)}_+\left(\frac{r_+}{r_-}\right)^{J^{(j)}_x}\right],
\end{equation}
where $E^{(j)}_+=\sum_{m=-j}^{j-1}|j,m+1;\vett k\>\<j,m;\vett
k|$. In Fig.~\ref{fig2} we report the plot of the scaling factor
$p(r)=r'/r$ for the phase covariant classical broadcasting procedure.
Notice that its performances are always better than in the universal
case reported in Fig.~\ref{f:fig1}. As expected, the single-site output
Bloch vector length~(\ref{eq:rphase}) coincides with the corresponding
quantity calculated for the optimal phase covariant superbroadcaster
in Ref.~\cite{Broad2} in the limit of infinite output copies.

Finally, notice that in the phase covariant case for states of the form
\ref{eq:phase-fam} the
NOT gate can always be achieved unitarily by a $\pi$-rotation around
the $z$ axis. Therefore the
optimal phase covariant NOT broadcasting has the same fidelity as the
optimal phase covariant superbroadcaster in Ref.~\cite{Broad2}.

\section{Asymptotic superbroadcasting and state estimation}
\label{sec:AsymptBroad&Est}

Recently, Bae and Ac\'in gave an argument to prove that the asymptotic cloning of pure states is
equivalent to state estimation \cite{acin}. The argument consists in noticing that, when restricted
to a single output Hilbert space, a symmetric cloning from $N$ to $M=\infty$ copies is an
entanglement breaking channel \cite{EBChannels}, and, therefore, it can be realized by the
semiclassical measure-and-prepare scheme, namely the single user output states are given by
\begin{equation}\label{MeasPrepOutput}
\rho_{out}^1 = \sum_i ~\Tr [P_i \rho^{\otimes N}] ~\rho_i~,
\end{equation} 
where the POVM $\{P_i\}$ represents the quantum measurement performed
on the input, and $\rho_i$ is the (single user) output state prepared
conditionally to the outcome $i$. As a consequence, if the input of
the cloning machine is the pure state $|\psi\>$, then the single site
cloning fidelity is $F_{clon}[N, \infty] = \<\psi|~\rho_{out}^1~
|\psi\>$, and coincides with the estimation fidelity $F_{est}[N] =
\sum_i~\Tr[P_i
\rho^{\otimes N}]~ \<\psi| \rho_i |\psi\>$ of the POVM $\{P_i\}$ with the guess states $\{\rho_i\}$.  This proves that the problem of optimal symmetric $N$-to-$\infty$ cloning is equivalent to the problem of optimal state estimation with $N$ input copies, and $F_{clon}[N, \infty]= F_{est}[N]$.

In the case of mixed states, a similar argument can be exploited to give a general explanation to
the fact that in the ideal case of infinite users the fidelity of the optimal superbroadcasting is
achieved by a semiclassical scheme. In fact, analogously to Ref. \cite{acin} since the output states
of superbroadcasting are invariant under permutations, for $M= \infty$ also the superbroadcasting
transformation is an entanglement breaking channel, when restricted to a single user. Therefore, it
can be realized by measurement and subsequent repreparation, and the single user output states are
written as in Eq. \eqref{MeasPrepOutput}, with suitable $\{P_i\}$ and $\{\rho_i\}$. Moreover, as for
the case of cloning, also in the case of superbroadcasting the figure of merit is the fidelity of
the output state with a \emph{pure} state---the eigenvector of the input density operator
corresponding to the maximal eigenvalue (see Eq. (\ref{Fid}) for the universal, and Eq.
(\ref{FidPhase}) for the phase covariant case).  It is then clear that asymptotically the fidelity of
the optimal universal (phase covariant) superbroadcasting coincides with that of the optimal
estimation of direction (phase). In general, the above reasoning shows that superbroadcasting with
infinite users is equivalent to the estimation of the eigenstate corresponding to the largest
eigenvalue of the input density matrix.  This result generalizes the well-known relation between
cloning and state estimation to the case of mixed states.

\section{Conclusions}
\label{sec:conc}

In this paper we considered the problem of quantum broadcasting, and in particular we analysed the
possibility of broadcasting $N$ input qubit states to $M$ output qubits with the same Bloch vector
direction, just by estimating the direction by a collective measurement on the input qubits and then
preparing $M$ outputs correspondingly. The main result is that this strategy allows to achieve
superbroadcasting, namely to have output copies which are even more pure than the input ones, at the
expense of classical correlations in the global output state. This superbroadcasting is suboptimal,
but asymptotically converges to the optimal one, confirming also in the case of mixed states the
fact that state estimation and cloning are asymptotically equivalent. We first considered the
universal broadcasting, and then the broadcasting of the antipodal state, the so called universal
NOT.  For this purpose, we proved that the semiclassical strategy is optimal.  Finally, we
considered the phase covariant version of the broadcasting problem, showing that superbroadcasting
occurs with suboptimal purification rate, which is still better than the one for universal
semiclassical superbroadcasting. The main interest of the summarized results is twofold. On one
hand, our results prove that superbroadcasting can be achieved by a semiclassical procedure, and
then coherent manipulation of quantum information is not necessary, even though optimal
superbroadcasting requires it. On the other hand, the practical interest of our results is that the
semiclassical rates exhibit a good approximation of the optimal rates, and can be more easily
achieved experimentally.

\acknowledgments This work has been supported in party by Ministero Italiano
dell'Universit\`a e della Ricerca (MIUR) through FIRB (bando 2001) and
PRIN 2005 and by the EC under the project SECOQC.


\begin{thebibliography}{50}
\bibitem{clon} W. K. Wootters and W. H.  Zurek, {\em Nature} {\bf
    299}, 802 (1982); D.~Dieks, Phys. Lett. A, {\bf 92}, 271 (1982);
  H. P. Yuen, Phys. Lett. A {\bf 113}, 405 (1986); G. C.  Ghirardi,
  referee report of N. Herbert, Found.  Phys.  {\bf 12}, 1171 (1982).
\bibitem{nobro} H. Barnum, C. M. Caves, C. A. Fuchs, R. Jozsa, and B.
  Schumacher, Phys. Rev. Lett. {\bf 76}, 2818 (1996).
\bibitem{Broad1} G. M. D'Ariano, C. Macchiavello, and P. Perinotti,
  Phys. Rev. Lett. {\bf 95}, 060503 (2005).
\bibitem{Broad2} F. Buscemi, G.M. D'Ariano, C. Macchiavello, and P.
  Perinotti, quant-ph/0602125
\bibitem{rafal} R. Demkowicz-Dobrzanski, Phys. Rev. A {\bf 71}, 062321
  (2005).
\bibitem{CEM} J. I. Cirac, A. K. Ekert, and C. Macchiavello, Phys.
  Rev. Lett. {\bf 82}, 4344 (1999).
\bibitem{Broad3} F. Buscemi, G.M. D'Ariano, C. Macchiavello, and P.
  Perinotti, quant-ph/0510155.
\bibitem{bruss-ek-macc} D.~Bru\ss{}, A.~Ekert, and C.~Macchiavello,
  Phys.~Rev.~Lett. {\bf 81}, 2598 (1998).
\bibitem{bruss-macc} D.~Bru\ss{} and C.~Macchiavello, Phys. Lett. A
  {\bf 253}, 249 (1999).
\bibitem{acin} J.~Bae and A.~Ac\'in,  Phys. Rev. Lett. {\bf 97}, 030402 (2006).
\bibitem{UNot} V.~Buzek, M.~Hillery, and R.~F.~Werner, Phys. Rev. A
  {\bf 60}, R2626 (1999).
\bibitem{Holevo} A. S.~Holevo, {\it Probabilistic and Statistical
    Aspects of Quantum Theory} (North Holland, Amsterdam 1982).
\bibitem{peris} F.~Buscemi, G.~M.~D'Ariano, P.~Perinotti, and
  M.~F.~Sacchi, Phys. Lett. A {\bf 314}, 374 (2003).
\bibitem{mixphase} G.~M.~D'Ariano, C.~Macchiavello, and P.~Perinotti,
  Phys. Rev. A {\bf 72}, 042327 (2005).
\bibitem{phasenot} F.~Buscemi, G.~M.~D'Ariano, and C.~Macchiavello,
  Phys. Rev. A {\bf 72}, 062311 (2005).
\bibitem{EBChannels}M. Horodecki, P. W. Shor and M. B. Ruskai, Rev. Math. Phys {\bf 15}, 629 (2003). 
\end{thebibliography}
\end{document}